\RequirePackage{lineno}
\documentclass[twocolumn,showpacs,byrevtex,reprint]{revtex4-1}

\usepackage{graphicx}
\usepackage{dcolumn}
\usepackage{bm}
\usepackage{rotating}
\usepackage{epstopdf}
\usepackage{color}
\usepackage{verbatim} 
\usepackage{multirow}
\usepackage[abs]{overpic}
\usepackage{amsmath}
\usepackage{url}

\newcommand{\PreserveBackslash}[1]{\let\temp=\\#1\let\\=\temp}
\newcolumntype{C}[1]{>{\PreserveBackslash\centering}p{#1}}
\newcolumntype{R}[1]{>{\PreserveBackslash\raggedleft}p{#1}}
\newcolumntype{L}[1]{>{\PreserveBackslash\raggedright}p{#1}}

\newcommand{\EE}{e^+e^-}

\newcommand{\too}{\rightarrow}

\begin{document}
\graphicspath{{figure/}}
\DeclareGraphicsExtensions{.eps,.png,.ps}

\title{\quad\\[0.0cm] \boldmath Study on the resonant parameters of $Y(4220)$ and $Y(4390)$ }

\author{Jielei Zhang}
\email{zhangjielei@ihep.ac.cn}
\author{Limin Yuan}
\author{Rumin Wang}
\affiliation{College of Physics and Electronic Engineering, Xinyang Normal University, Xinyang 464000, People's Republic of China}

\begin{abstract}
Many vector charmonium-like states have been reported recently in the cross sections of $\EE \too \omega\chi_{c0}$, $\pi^{+}\pi^{-}h_c$, $\pi^{+}\pi^{-}J/\psi$, $\pi^{+}\pi^{-}\psi(3686)$ and $\pi^{+}D^{0}D^{*-}+c.c.$ To better understand the nature of these states, a combined fit is performed to these cross sections by using three resonances $Y(4220)$, $Y(4390)$ and $Y(4660)$. The resonant parameters for the three resonances are obtained. We emphasize that two resonances $Y(4220)$ and $Y(4390)$ are sufficient to explain these cross sections below 4.6 GeV. The lower limits of $Y(4220)$ and $Y(4390)$'s leptonic decay widths are also determined to be $(36.4\pm2.0\pm4.2)$ and $(123.8\pm6.5\pm9.0)$ eV.
\end{abstract}

\maketitle

In the last decade, charmonium physics has gained renewed strong interest
from both the theoretical and the experimental side, due to the
observation of a series of charmonium-like states, such as the
$X(3872)$~\cite{X3872}, the $Y(4260)$~\cite{Y4260-babar} and the
$Y(4360)$~\cite{Y4360-babar}.
These states do not fit in the conventional level system of charmonium states
and are good candidates for exotic states not encompassed by the naive quark
model~\cite{Y-theory1}. Moreover, many charged charmonium-like states or their neutral
partners~\cite{Y-theory4} were observed, which might indicate the presence of new dynamics in this energy region.

$Y(4260)$ is the first charmonium-like state, which was observed in the process $\EE \too \pi^+\pi^-J/\psi$ by the $BABAR$ experiment using an initial-state-radiation (ISR) technique~\cite{Y4260-babar}. This observation was immediately confirmed by the CLEO~\cite{Y4260-cleo} and Belle experiments~\cite{Y4260-belle} in the same process. Being produced in $\EE$ annihilation, the $Y$ state has quantum numbers $J^{PC}=1^{--}$. $Y(4360)$ is the second $Y$ state, which was observed in the $\EE \too \gamma_{\text{ISR}}Y(4360) \too \gamma_{\text{ISR}}\pi^+\pi^-\psi(3686)$ by $BABAR$~\cite{Y4360-babar} and subsequently confirmed by Belle experiment~\cite{Y4360-belle}. Belle also observed another structure, $Y(4660)$, in the $\pi^+\pi^-\psi(3686)$~\cite{Y4360-belle}. The observation of these $Y$ states has stimulated substantial theoretical discussions on their nature~\cite{Y-theory1}.

Recently, with higher statistic data, the $\EE \too \pi^+\pi^-J/\psi$ cross section was measured by BESIII experiment more precisely~\cite{pipijpsi-bes}. The fine structure was observed for $Y(4260)$ in $\EE \too \pi^+\pi^-J/\psi$. The $Y(4260)$ structure is a combination of two resonances, the lower one is $Y(4220)$ and the higher is $Y(4320)$. Using the results for $\EE \too \pi^+\pi^-\psi(3686)$ from Belle~\cite{pipipsip-belle}, $BABAR$~\cite{pipipsip-babar} and BESIII experiments~\cite{pipipsip-bes}, the authors of Ref.~\cite{zhang} also observed the fine structure for $Y(4360)$ in $\EE \too \pi^+\pi^-\psi(3686)$, inferring that the $Y(4360)$ structure is also a combination of two resonances, the lower one is $Y(4220)$ and the higher is $Y(4360)$. The $Y(4220)$ state also is observed in the processes $\EE \too \omega\chi_{c0}$~\cite{omegachic, omegachic2}, $\pi^+\pi^-h_c$~\cite{pipihc-bes} and $\pi^{+}D^{0}D^{*-}+c.c.$~\cite{piDDstar} by BESIII experiment. In the $\EE \too \pi^+\pi^-h_c$ and $\pi^{+}D^{0}D^{*-}+c.c.$, besides the $Y(4220)$, another $Y$ state $Y(4390)$ is observed~\cite{pipihc-bes, piDDstar}. The parameters for $Y(4220)$, $Y(4320)$, $Y(4360)$ and $Y(4390)$ states in different processes are listed in Table~\ref{tab:parameter}. In addition, Authors of Ref.~\cite{shen} have performed a combine fit to the cross sections of $\EE \too \omega\chi_{c0}$, $\pi^{+}\pi^{-}h_c$, $\pi^{+}\pi^{-}J/\psi$ and $\pi^{+}D^{0}D^{*-}+c.c.$ to obtain the resonant parameters for $Y(4220)$, $Y(4320)$ and $Y(4390)$ states.

\begin{table*}[htbp]
\begin{center}
\caption{ The parameters for $Y(4220)$ ($\omega\chi_{c0}$, $\pi^{+}\pi^{-}h_c$, $\pi^{+}\pi^{-}J/\psi$, $\pi^{+}\pi^{-}\psi(3686)$ and $\pi^{+}D^{0}D^{*-}+c.c.$ ), $Y(4320)$ ($\pi^{+}\pi^{-}J/\psi$), $Y(4360)$ ($\pi^{+}\pi^{-}\psi(3686)$) and $Y(4390)$ ($\pi^{+}\pi^{-}h_c$ and $\pi^{+}D^{0}D^{*-}+c.c.$) states in different processes.  The first uncertainties are statistical, and the second systematic.}
\label{tab:parameter}
\begin{tabular}{ccccc}
  \hline
  \hline
   & \multicolumn{2}{c}{~~~~~~~~~$Y(4220)$~~~~~~~} & \multicolumn{2}{c}{~~~~~~$Y(4320)/Y(4360)/Y(4390)$~~~}  \\
   & ~~~~~~$M$ (MeV/$c^2$)~~~~~ & ~~~~~~~$\Gamma$ (MeV)~~~~~~~~ & ~~~~~~~$M$ (MeV/$c^2$)~~~~~ & ~~~$\Gamma$ (MeV)~~~  \\
  \hline
  $\omega\chi_{c0}$~\cite{omegachic2} & $4226\pm8\pm6$ & $39\pm12\pm2$ & &   \\
  $\pi^+\pi^-h_c$~\cite{pipihc-bes} & $4218.4^{+5.5}_{-4.5}\pm0.9$ & $66.0^{+12.3}_{-8.3}\pm0.4$ & $4391.5^{+6.3}_{-6.8}\pm1.0$ & $139.5^{+16.2}_{-20.6}\pm0.6$   \\
  $\pi^+\pi^-J/\psi$~\cite{pipijpsi-bes} & $4222.0\pm3.1\pm1.4$ & $44.1\pm4.3\pm2.0$ & $4320.0\pm10.4\pm7.0$ & $101.4^{+25.3}_{-19.7}\pm10.2$   \\
  $\pi^+\pi^-\psi(3686)$~\cite{zhang} & $4209.1\pm6.8\pm7.0$ & $76.6\pm14.2\pm2.4$ & $4383.7\pm2.9\pm6.2$ & $94.2\pm7.3\pm2.0$   \\
  $\pi^{+}D^{0}D^{*-}+c.c.$~\cite{piDDstar} & $4224.8\pm5.6\pm4.0$ & $72.3\pm9.1\pm0.9$ & $4400.1\pm9.3\pm2.1$ & $181.7\pm16.9\pm7.4$   \\
  \hline
  \hline
\end{tabular}
\end{center}
\end{table*}

These states challenge the understanding of charmonium spectroscopy as well as QCD calculations~\cite{Y-theory1, Y-theory2, Y-theory3}. According to potential models, there are five vector charmonium states between the 1$D$ state $\psi(3770)$ and 4.7 GeV/$c^2$, namely, the 3$S$, 2$D$, 4$S$, 3$D$ and 5$S$ states~\cite{Y-theory1}. Besides the three well-established structures observed in the inclusive hadronic cross section~\cite{pdg}, i.e., $\psi(4040)$, $\psi(4160)$ and $\psi(4415)$, five $Y$ states, i.e., $Y(4220)$, $Y(4320)$, $Y(4360)$, $Y(4390)$ and $Y(4660)$ have been observed. These newly-observed $Y$ states exceed the number of vector charmonium states predicted by potential models in this energy region. They are thus good candidates for exotic states, such as hybrid states, tetraquark states and molecule states~\cite{Y-theory4}.

\begin{figure}[htbp]
\begin{center}
\includegraphics[width=0.23\textwidth]{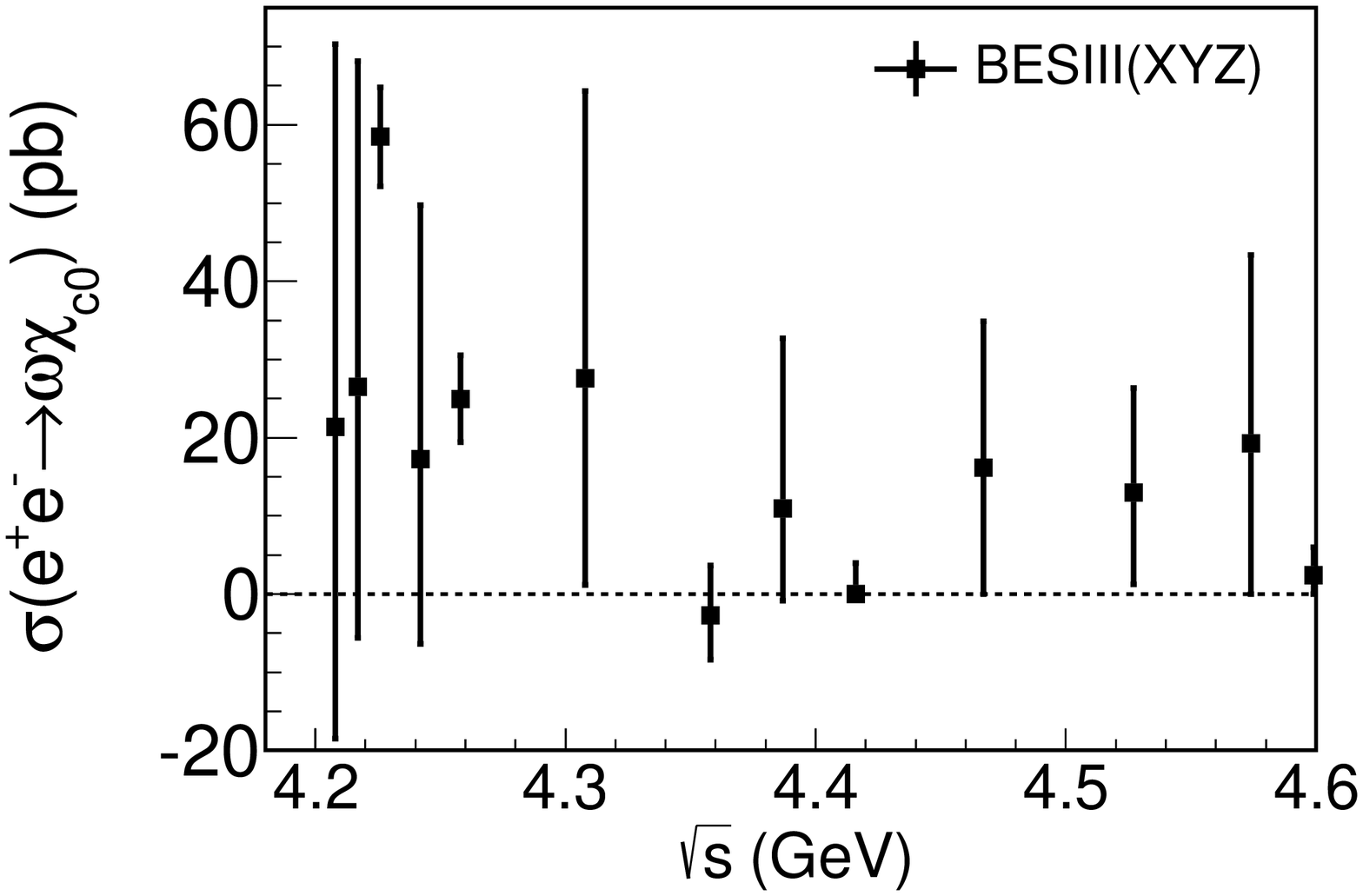}
\includegraphics[width=0.23\textwidth]{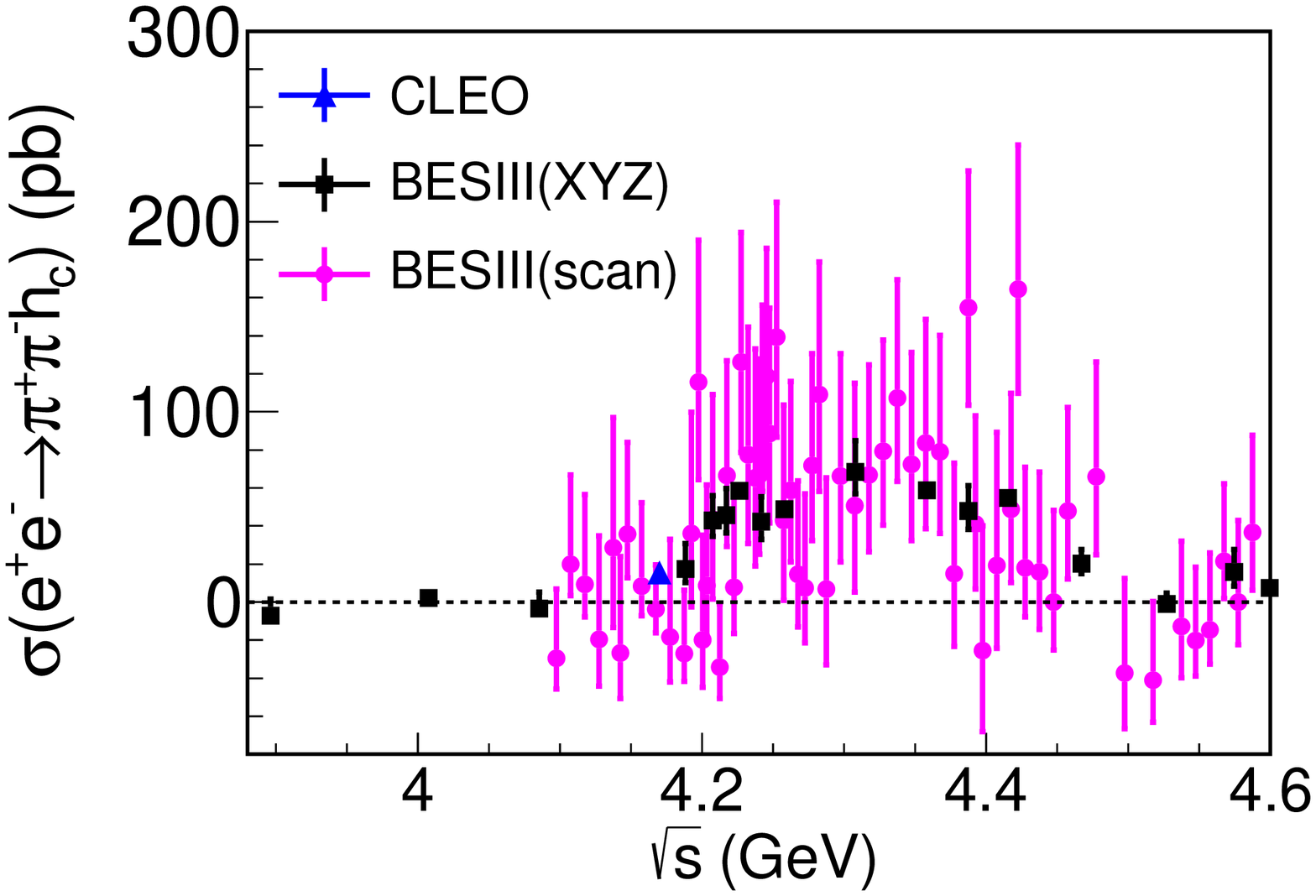}
\includegraphics[width=0.23\textwidth]{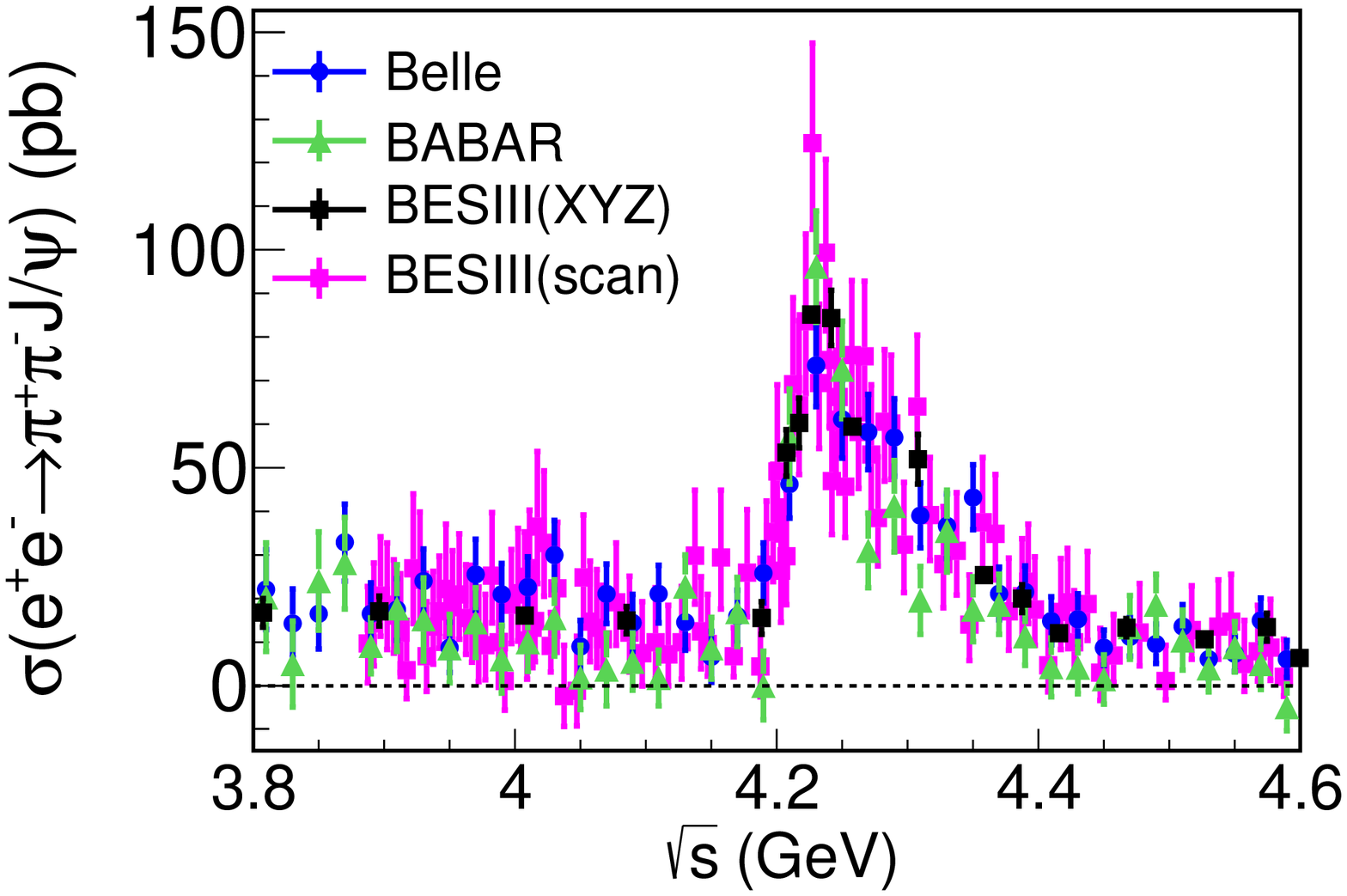}
\includegraphics[width=0.23\textwidth]{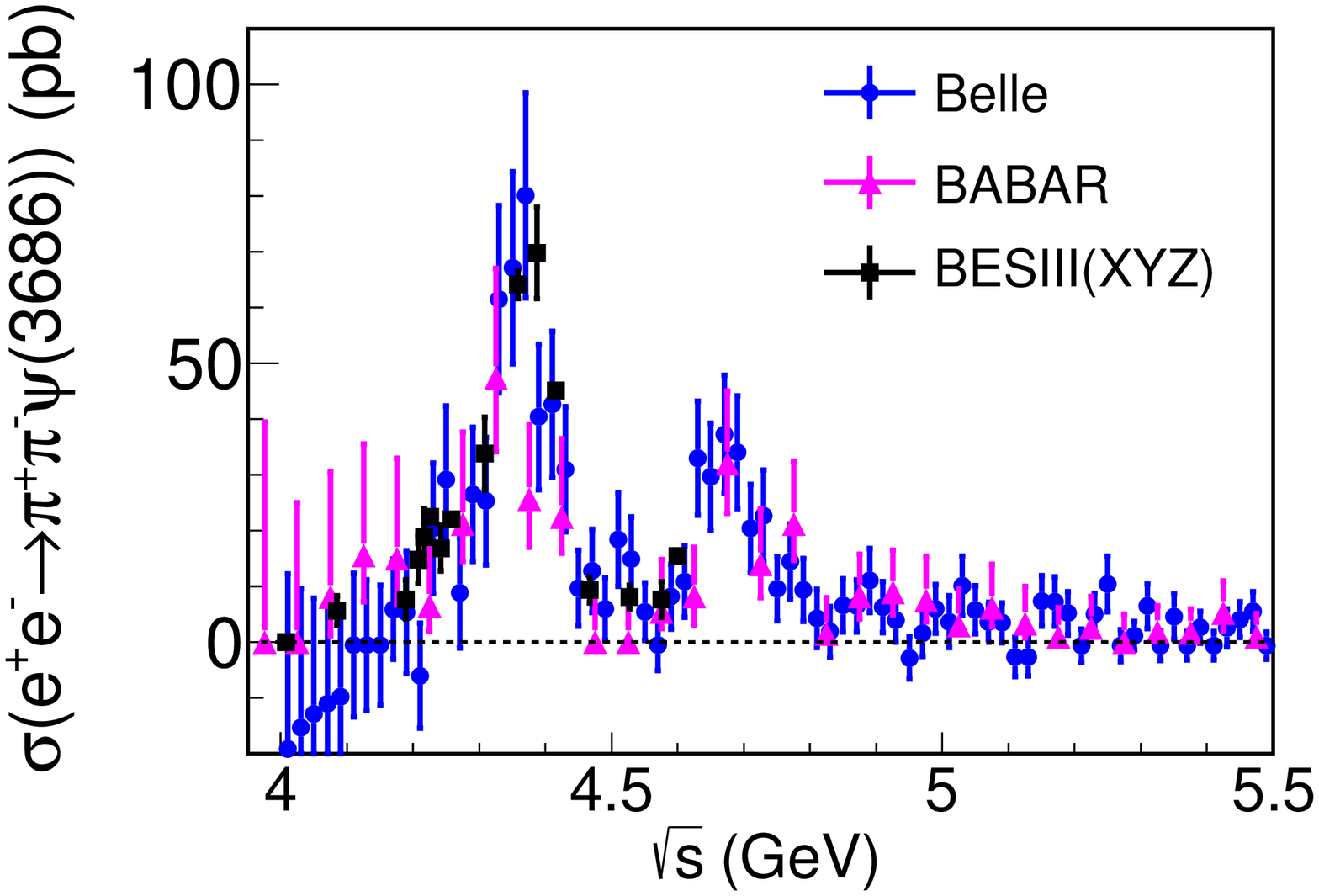}
\includegraphics[width=0.23\textwidth]{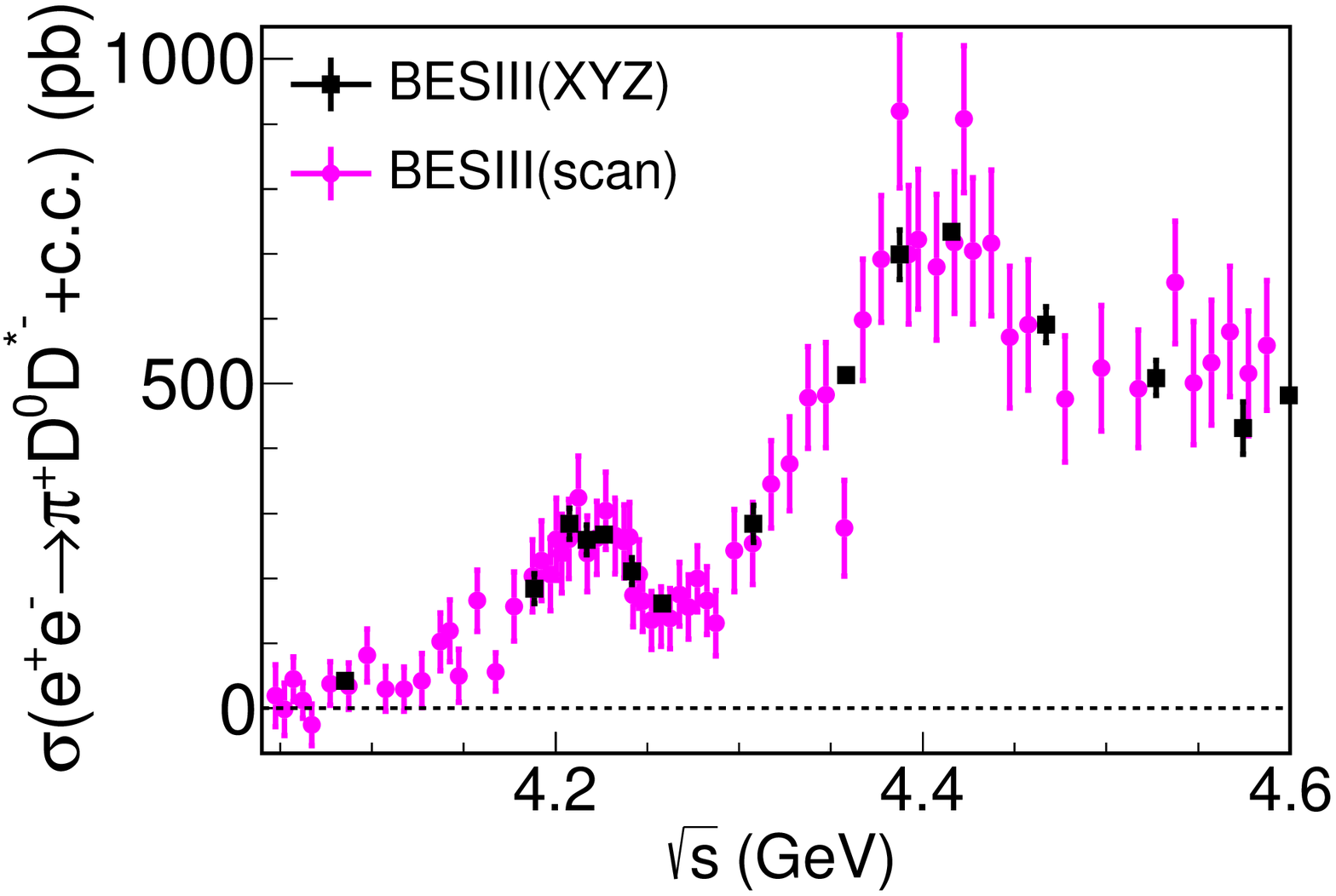}
\includegraphics[width=0.23\textwidth]{empty.eps}
\caption{Cross sections of $\EE \too \omega\chi_{c0}$, $\pi^{+}\pi^{-}h_c$, $\pi^{+}\pi^{-}J/\psi$, $\pi^{+}\pi^{-}\psi(3686)$ and $\pi^{+}D^{0}D^{*-}+c.c.$ measured by Belle, $BABAR$, CLEO and BESIII experiments.}
\label{fig:crosssection}
\end{center}
\end{figure}

Figure~\ref{fig:crosssection} shows the cross sections of $\EE \too \omega\chi_{c0}$~\cite{omegachic, omegachic2}, $\pi^{+}\pi^{-}h_c$~\cite{pipihc-bes, pipihc-cleo}, $\pi^{+}\pi^{-}J/\psi$~\cite{pipijpsi-bes, pipijpsi-belle, pipijpsi-babar}, $\pi^{+}\pi^{-}\psi(3686)$~\cite{pipipsip-belle, pipipsip-babar, pipipsip-bes} and $\pi^{+}D^{0}D^{*-}+c.c.$~\cite{piDDstar} measured by Belle, $BABAR$, CLEO and BESIII experiments. For data from BESIII, ``XYZ" data sample refers to the energy points with integrated luminosity larger than 40 pb$^{-1}$ and ``scan" data sample refers to the energy points with integrated luminosity smaller than 20 pb$^{-1}$. In this paper, we perform a combined fit to these cross sections.

These vector charmonium-like states in the fit are assumed to be resonances. We parameterize the cross section with the coherent sum of a few amplitudes, either resonance represented by a Breit-Wigner (BW) function or non-resonant production term parameterized with a phase space function or an exponential function. The BW function used in this article is~\cite{shen}

\begin{equation}
\begin{aligned}
BW(\sqrt{s})=\frac{\sqrt{12\pi\Gamma_{e^+e^-}\mathcal{B}_{f}\Gamma}}{s-M^{2}+iM\Gamma}\sqrt{\frac{PS(\sqrt{s})}{PS(M)}},
\end{aligned}
\end{equation}
where $M$ and $\Gamma$ are the mass and total width of the resonance, respectively; $\Gamma_{e^+e^-}$ is the partial width to $e^+e^-$; $\mathcal{B}_{f}$ is the branching fraction of the resonance decays into final state $f$, and $PS(\sqrt{s})$ is the phase space factor that increases smoothly from the mass threshold with the $\sqrt{s}$~\cite{pdg}. In the fit, the $\Gamma_{e^+e^-}$ and the $\mathcal{B}_{f}$ can not be obtained separately, we can only extract the product $\Gamma_{e^+e^-}\mathcal{B}_f$.

Ref.~\cite{shen} has performed a combine fit to the cross sections of $\EE \too \omega\chi_{c0}$, $\pi^{+}\pi^{-}h_c$, $\pi^{+}\pi^{-}J/\psi$ and $\pi^{+}D^{0}D^{*-}+c.c.$, while the cross section of $\EE \too \pi^{+}\pi^{-}\psi(3686)$ is not included. In Ref.~\cite{shen}, the resonances $Y(4320)$ and $Y(4390)$ are regarded as different states in the fit, while from Table~\ref{tab:parameter}, we notice that the parameters for $Y(4320)$, $Y(4360)$ and $Y(4390)$ are relatively close. Although there are some differences in the obtained mass and width in different channels, it may due to there are only a few data points with small errors around 4.4 GeV. It is not reasonable that there are three states in such a close position. In addition, the analysis in Ref.~\cite{pipijpsi} also indicates that the charmonium-like states $Y(4360)$ in the $\pi^+\pi^-\psi(3686)$ and $Y(4320)$ in the $\pi^+\pi^-J/\psi$ should be the same state. Therefore, we consider $Y(4320)$, $Y(4360)$ and $Y(4390)$ as the same state, which has been suggested in Ref.~\cite{soto}. The same state is marked as ``$Y(4390)$" in this paper.
A least $\chi^{2}$ fit method is used to perform a combined fit to the five cross sections using three resonances $Y(4220)$, $Y(4390)$ and $Y(4660)$, assuming the two resonances $Y(4220)$ and $Y(4390)$ are the same two states in these processes. The fit functions are,

\begin{equation}
\begin{aligned}
\sigma_{\omega\chi_{c0}}(\sqrt{s})= & |BW_{1}(\sqrt{s})|^{2},
\end{aligned}
\end{equation}
\begin{equation}
\begin{aligned}
\sigma_{\pi^+\pi^-h_c}(\sqrt{s})= & |BW_{1}(\sqrt{s})+BW_{2}(\sqrt{s})e^{i\phi_{1}}|^{2},
\end{aligned}
\end{equation}
\begin{equation}
\begin{aligned}
\sigma_{\pi^+\pi^-J/\psi}(\sqrt{s})= & |c_1\sqrt{EXP(\sqrt{s})}+BW_{1}(\sqrt{s})e^{i\phi_{2}} \\
& +BW_{2}(\sqrt{s})e^{i\phi_{3}}|^{2},
\end{aligned}
\end{equation}
\begin{equation}
\begin{aligned}
\sigma_{\pi^+\pi^-\psi(3686)}(\sqrt{s})= & |BW_{1}(\sqrt{s})+BW_{2}(\sqrt{s})e^{i\phi_{4}} \\
& +BW_{3}(\sqrt{s})e^{i\phi_{5}}|^{2},
\end{aligned}
\end{equation}
\begin{equation}
\begin{aligned}
\sigma_{\pi^+D^0D^{*-}+c.c.}(\sqrt{s})= & |c_2\sqrt{PS(\sqrt{s})}+BW_{1}(\sqrt{s})e^{i\phi_{6}} \\
& +BW_{2}(\sqrt{s})e^{i\phi_{7}}|^{2},
\end{aligned}
\end{equation}
where $BW_{1}$, $BW_{2}$ and $BW_{3}$ denote the resonances $Y(4220)$, $Y(4390)$ and $Y(4660)$, respectively; $PS(\sqrt{s})$ is the phase space factor; $EXP(\sqrt{s})=e^{-p_0(\sqrt{s}-M_{th})}PS(\sqrt{s})$, is an exponential function, where $p_0$ is free parameter, $M_{th}=2m_{\pi}+m_{J/\psi}$ is the mass threshold of the $\pi^+\pi^-J/\psi$ system; $\phi_{1}$, $\phi_{2}$, $\phi_{3}$, $\phi_{4}$, $\phi_{5}$, $\phi_{6}$ and $\phi_{7}$ are relative phases; $c_1$ and $c_2$ are amplitudes of exponential function term and phase space term.

\begin{figure*}[htbp]
\begin{center}
\includegraphics[width=0.23\textwidth]{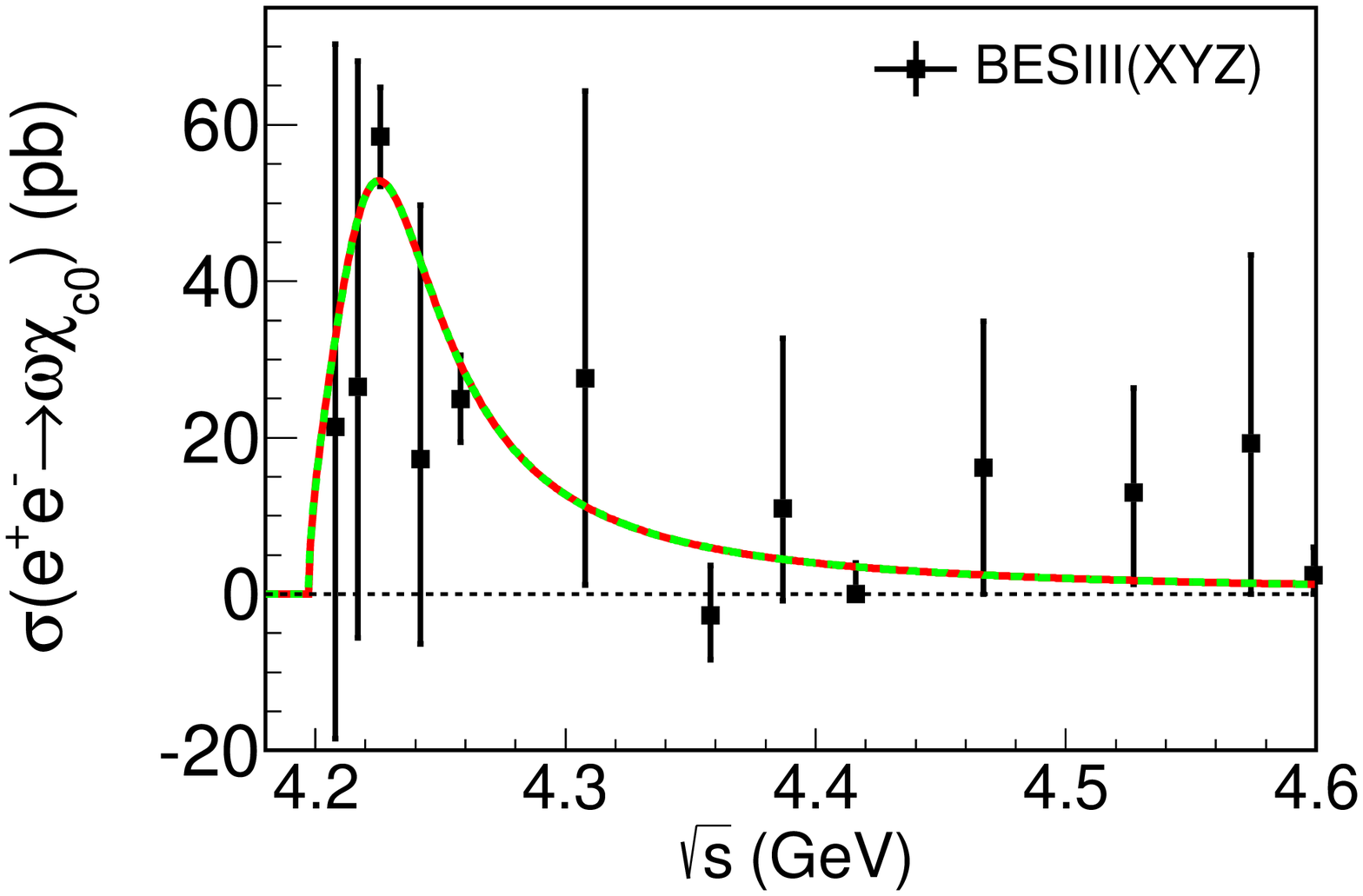}
\includegraphics[width=0.23\textwidth]{empty.eps}
\includegraphics[width=0.23\textwidth]{empty.eps}
\includegraphics[width=0.23\textwidth]{empty.eps}
\includegraphics[width=0.23\textwidth]{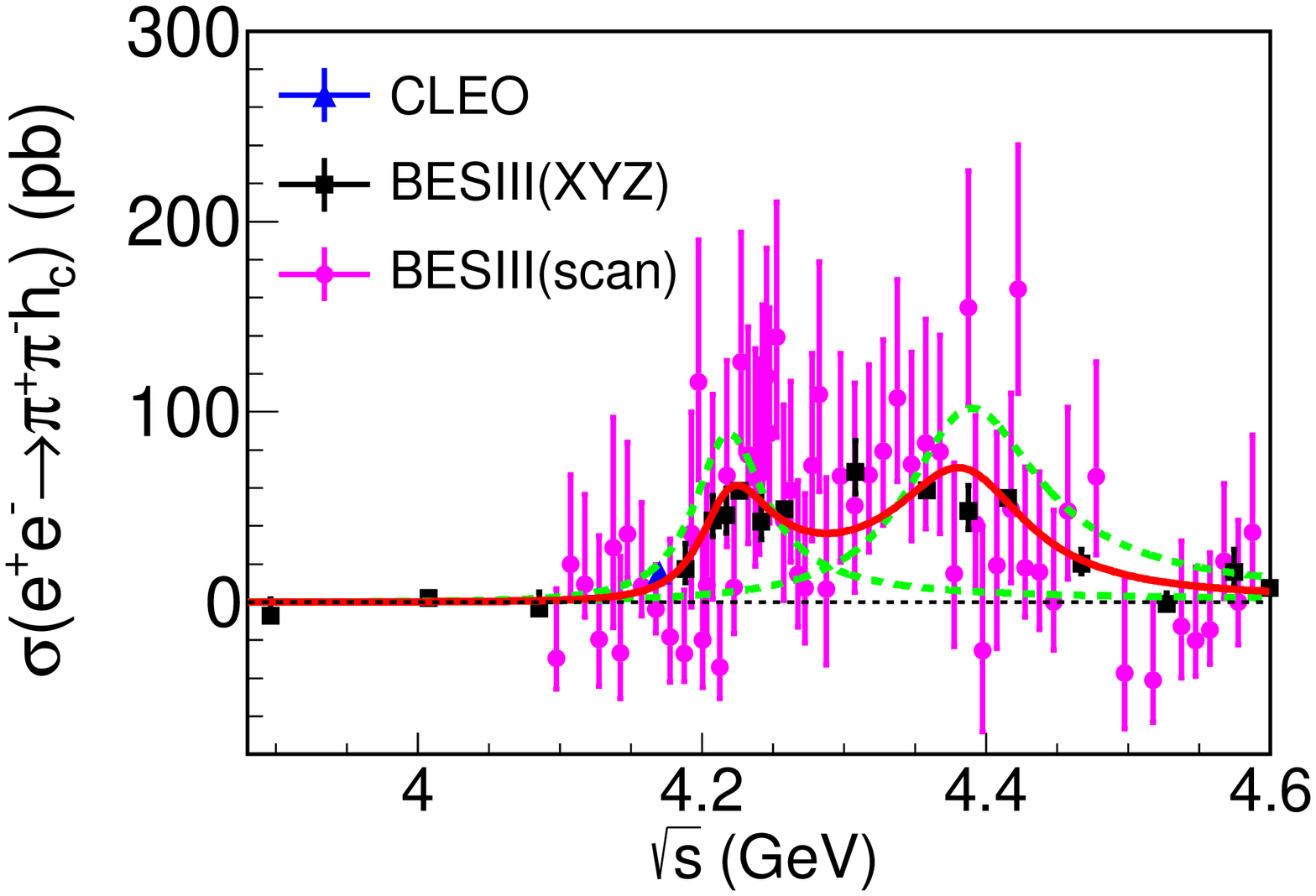}
\includegraphics[width=0.23\textwidth]{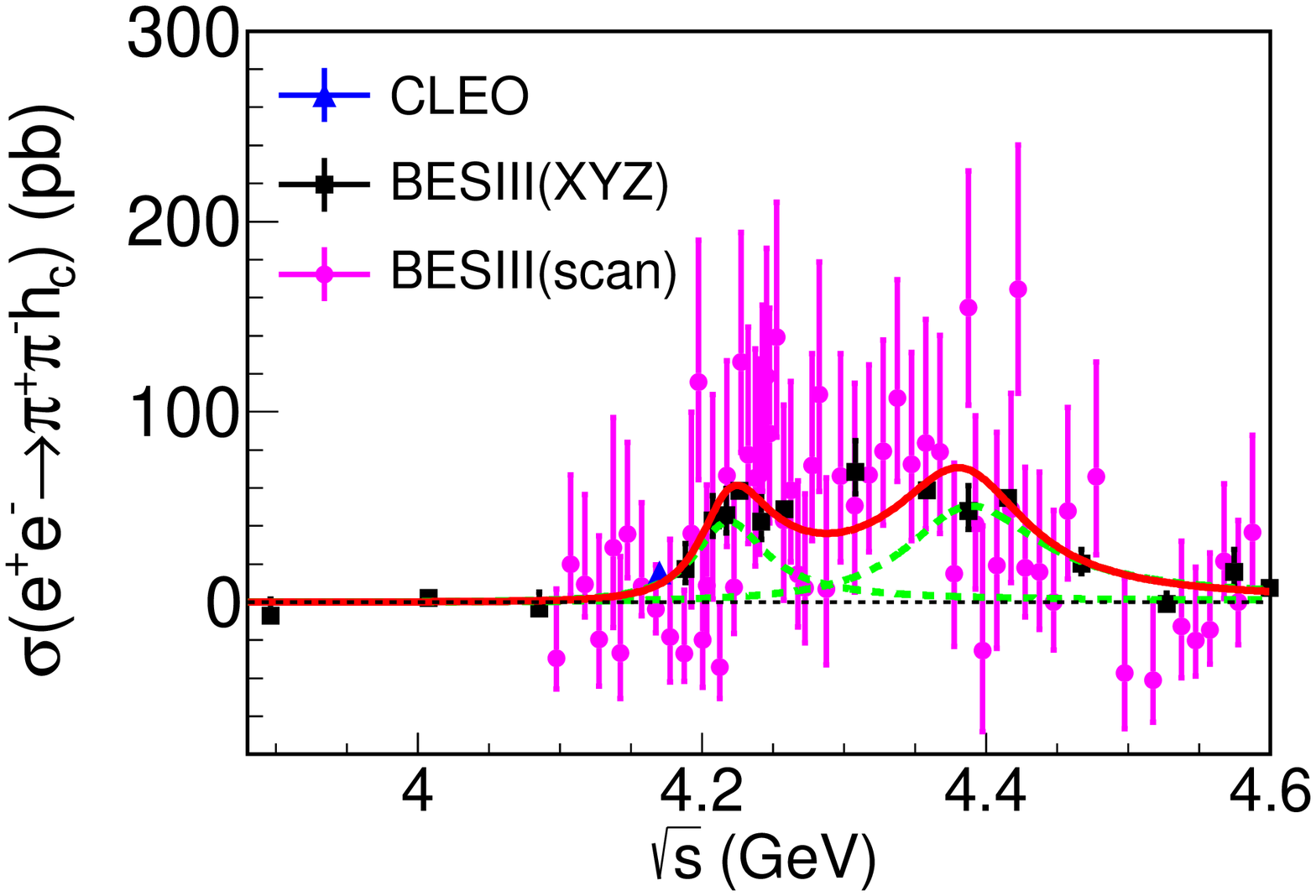}
\includegraphics[width=0.23\textwidth]{empty.eps}
\includegraphics[width=0.23\textwidth]{empty.eps}
\includegraphics[width=0.23\textwidth]{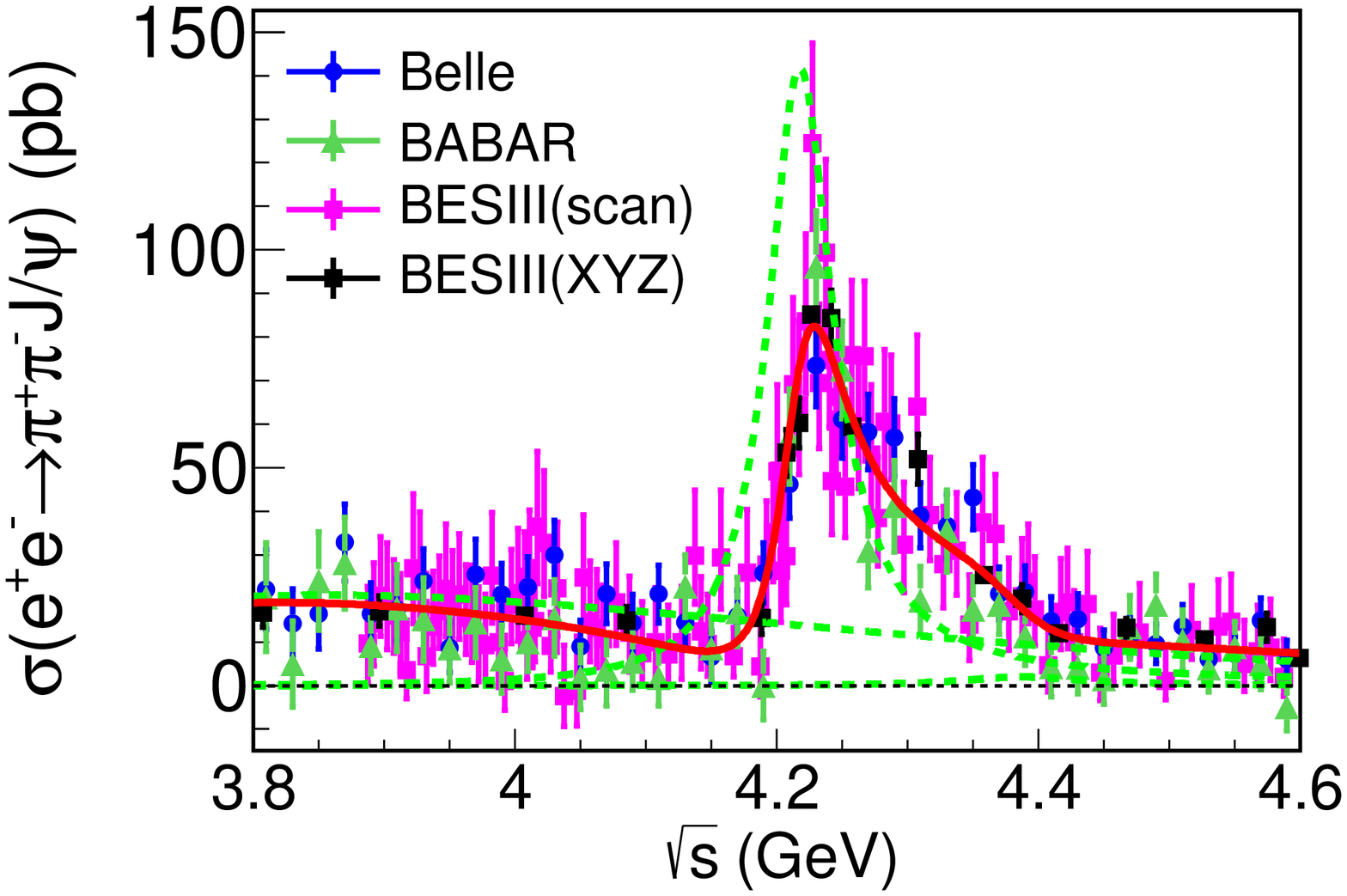}
\includegraphics[width=0.23\textwidth]{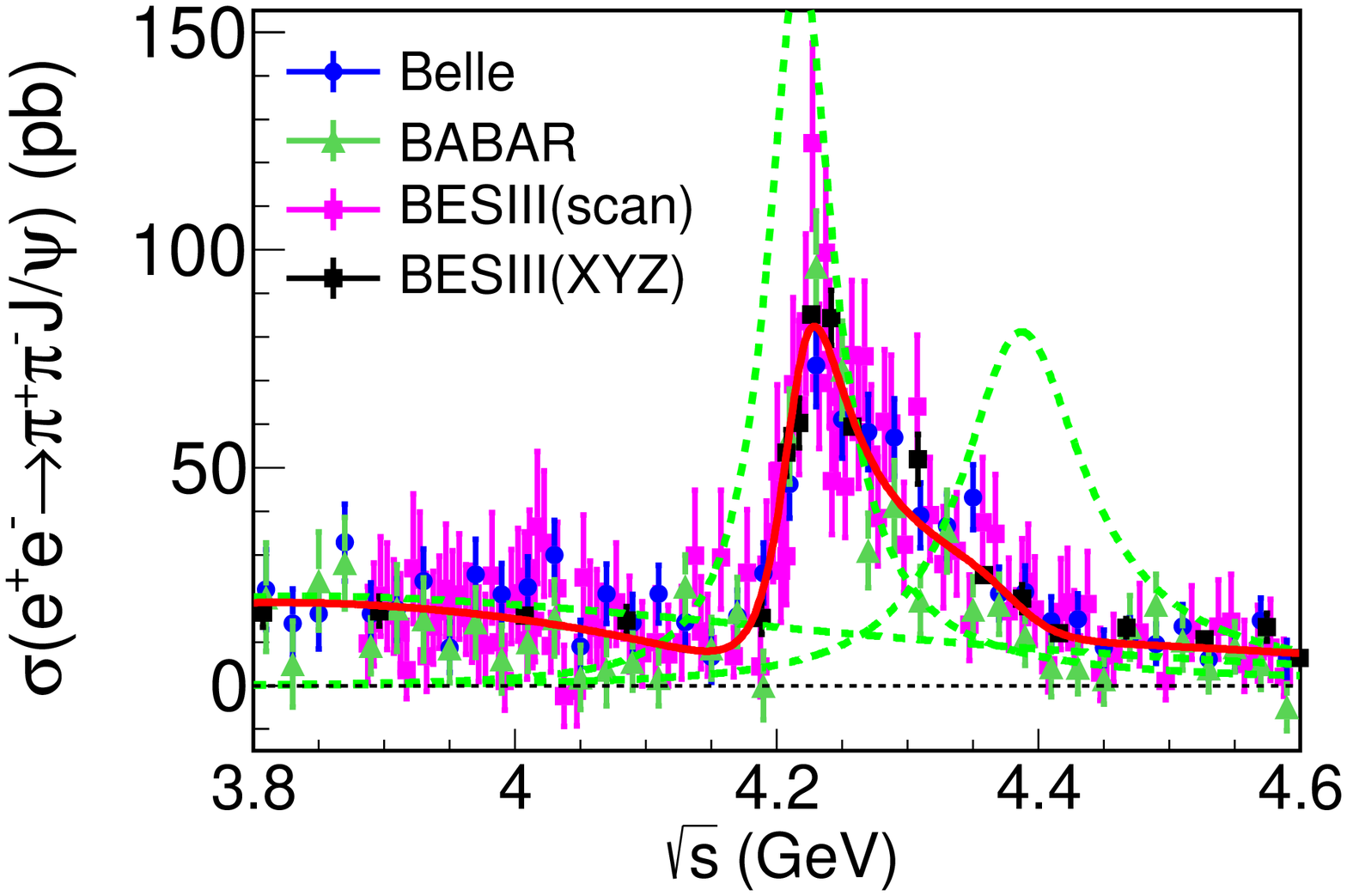}
\includegraphics[width=0.23\textwidth]{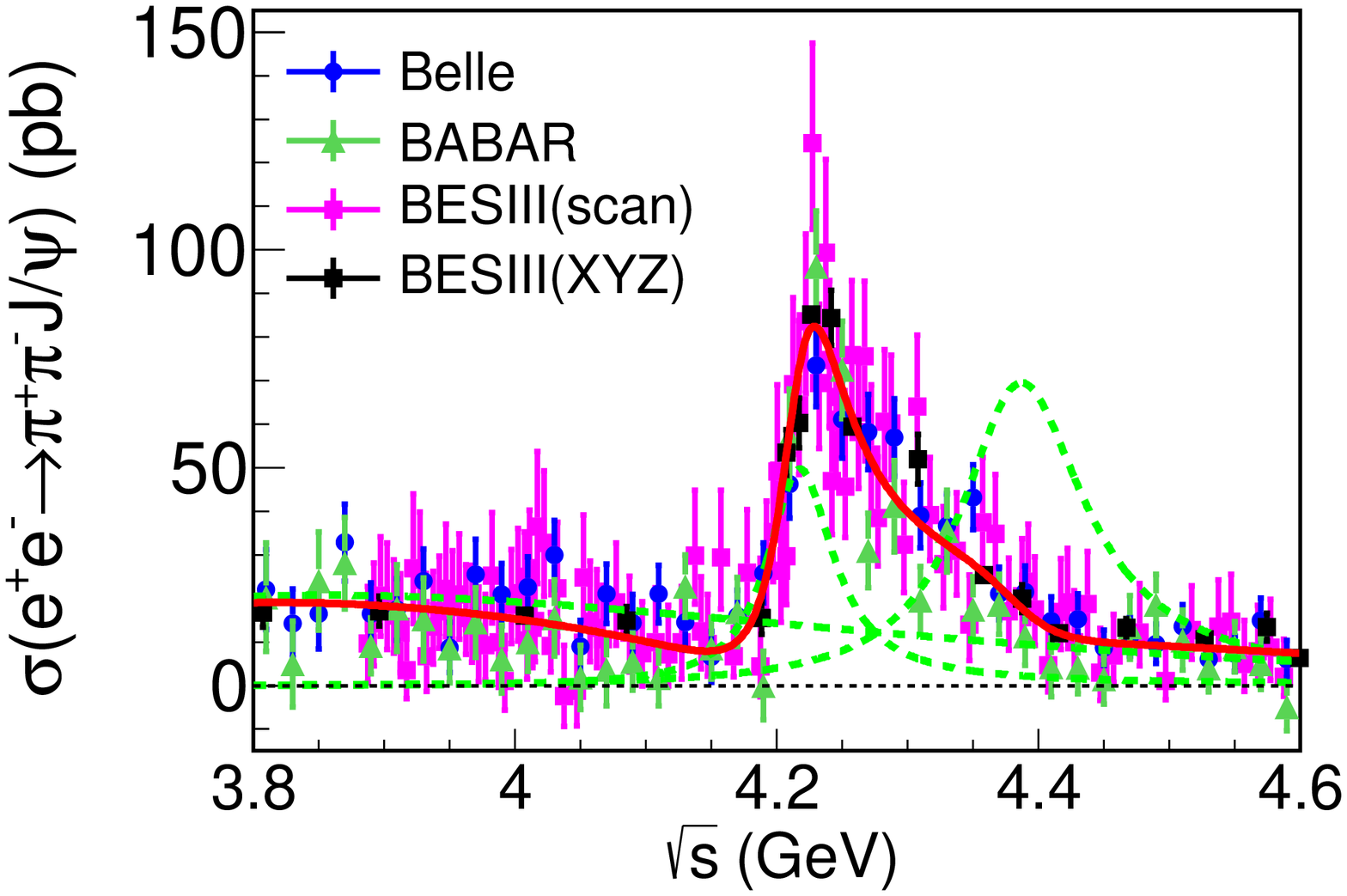}
\includegraphics[width=0.23\textwidth]{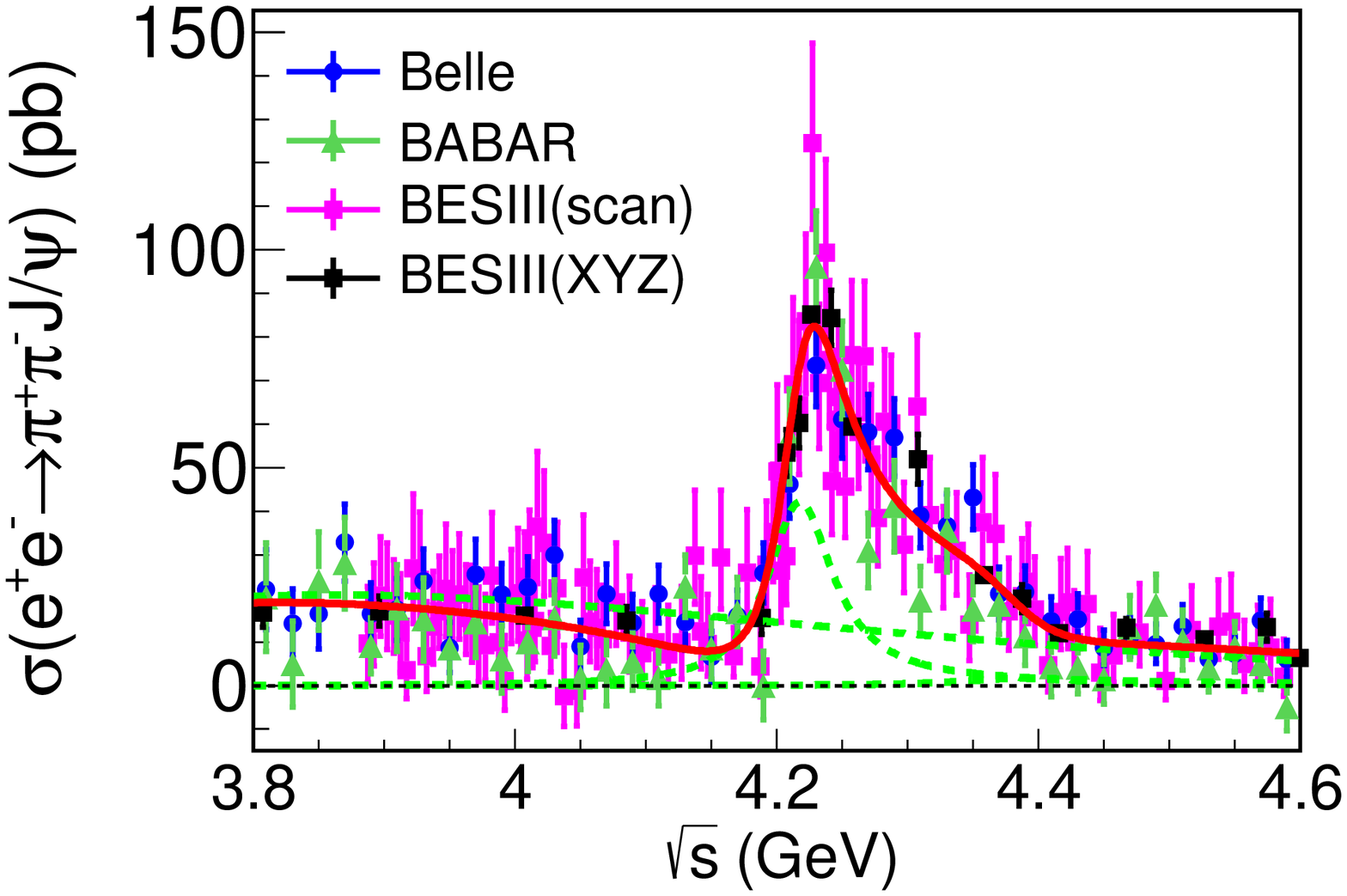}
\includegraphics[width=0.23\textwidth]{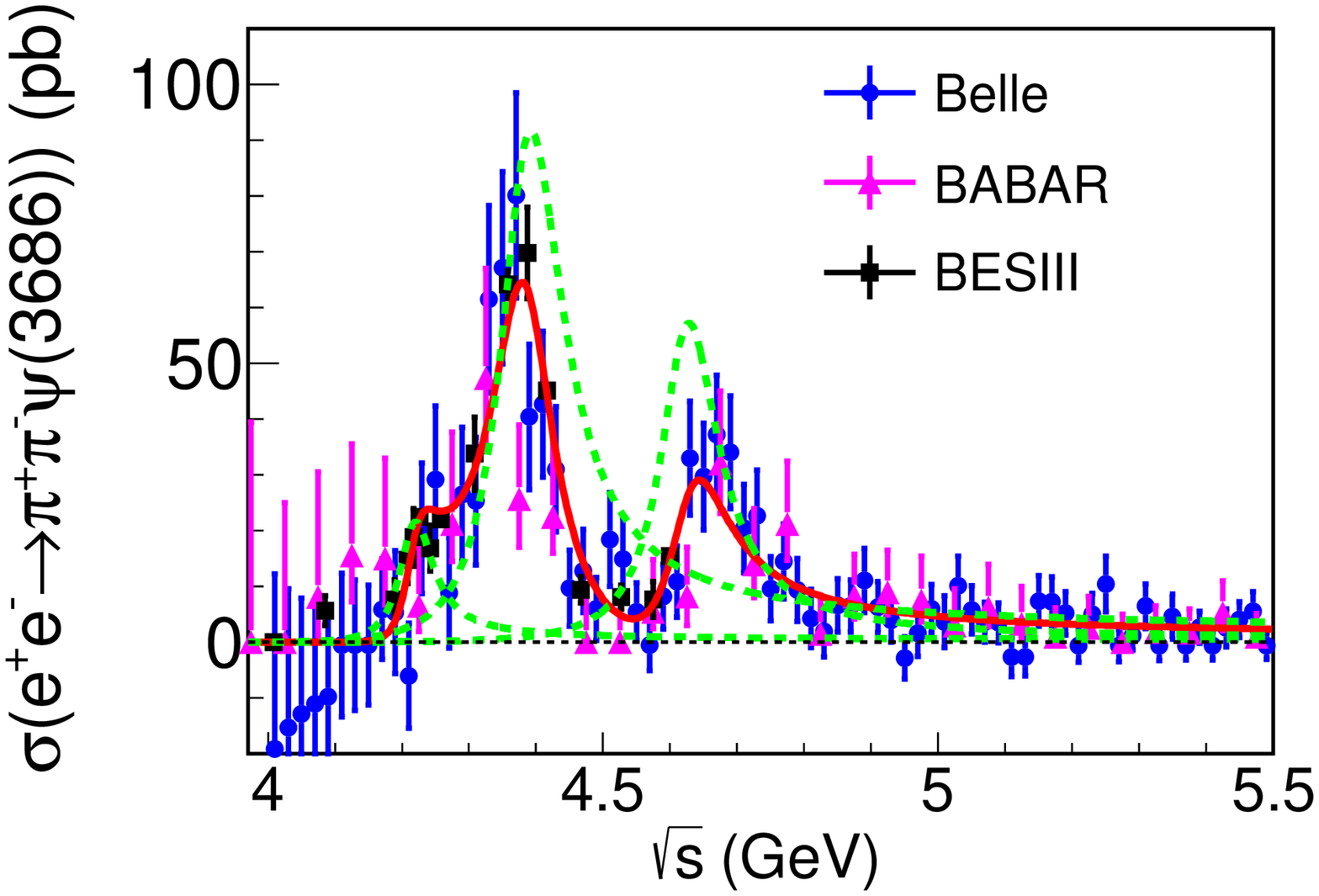}
\includegraphics[width=0.23\textwidth]{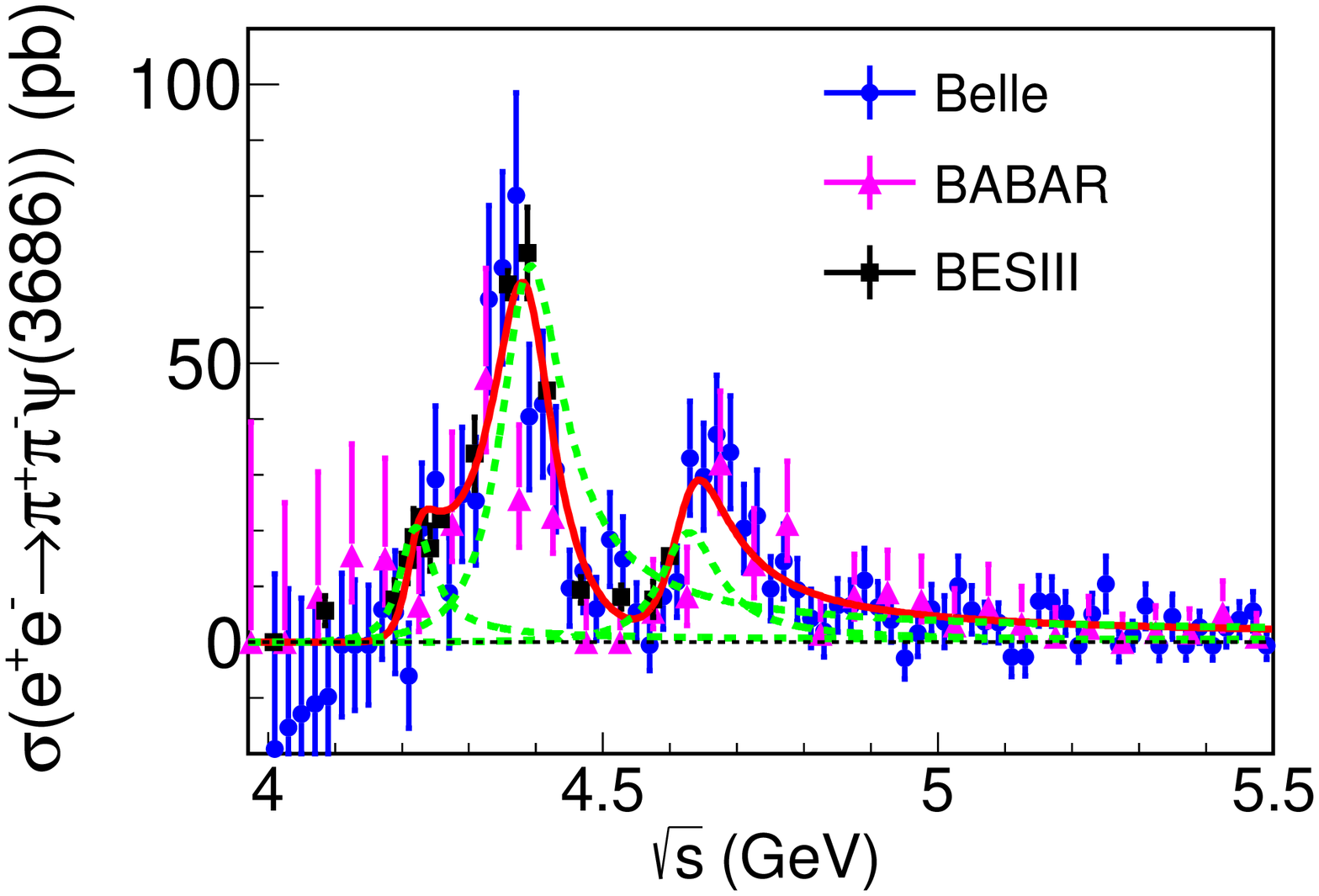}
\includegraphics[width=0.23\textwidth]{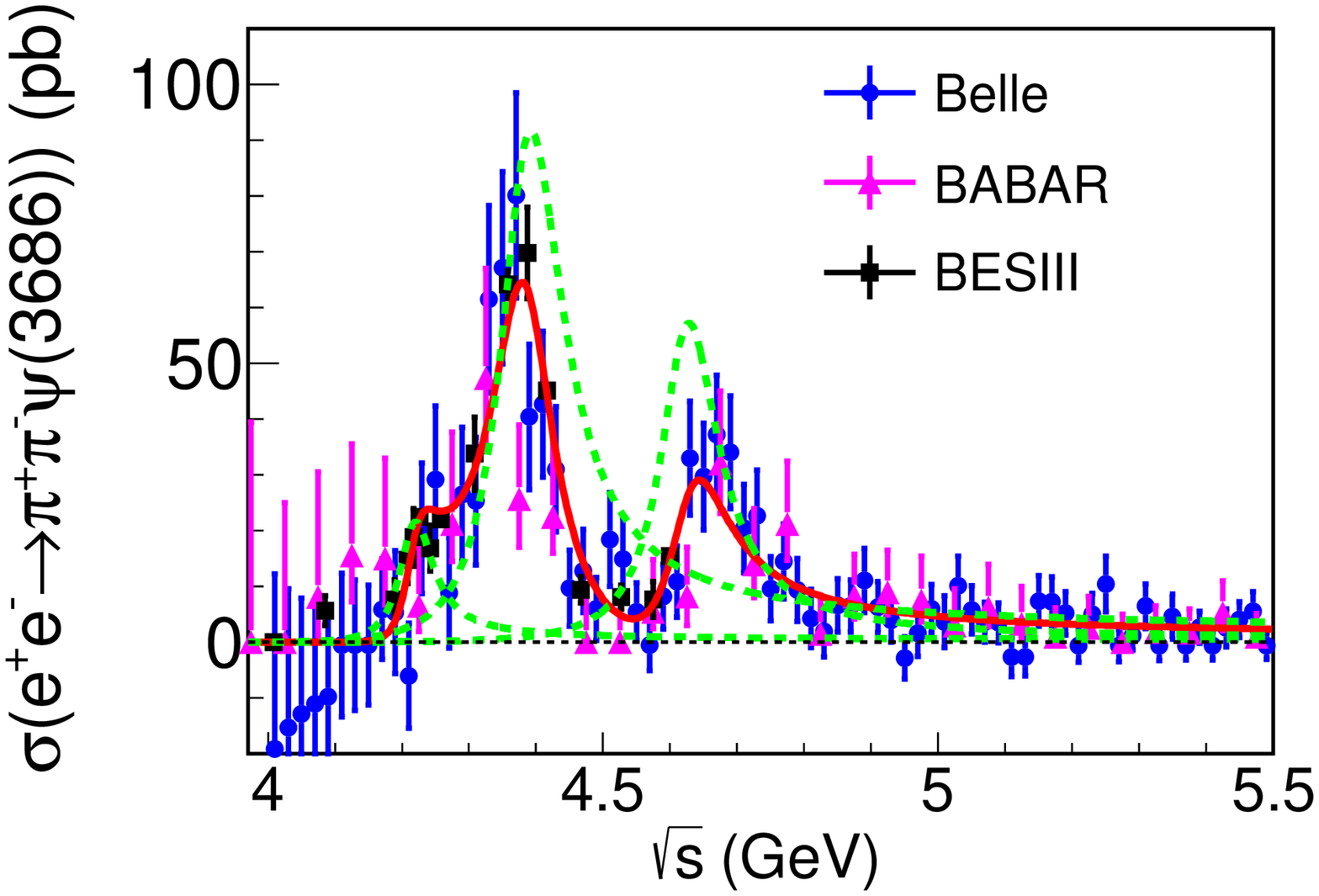}
\includegraphics[width=0.23\textwidth]{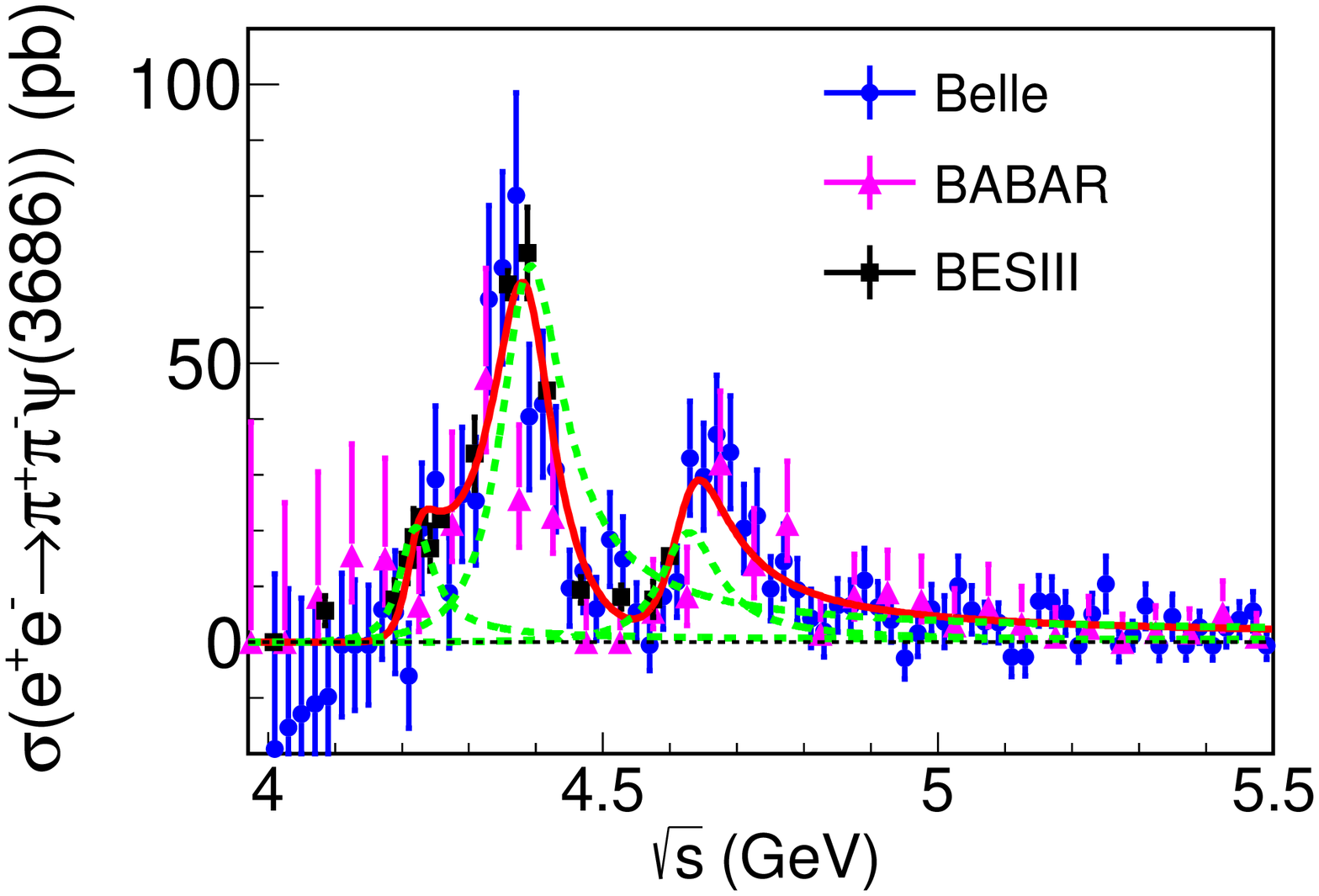}
\includegraphics[width=0.23\textwidth]{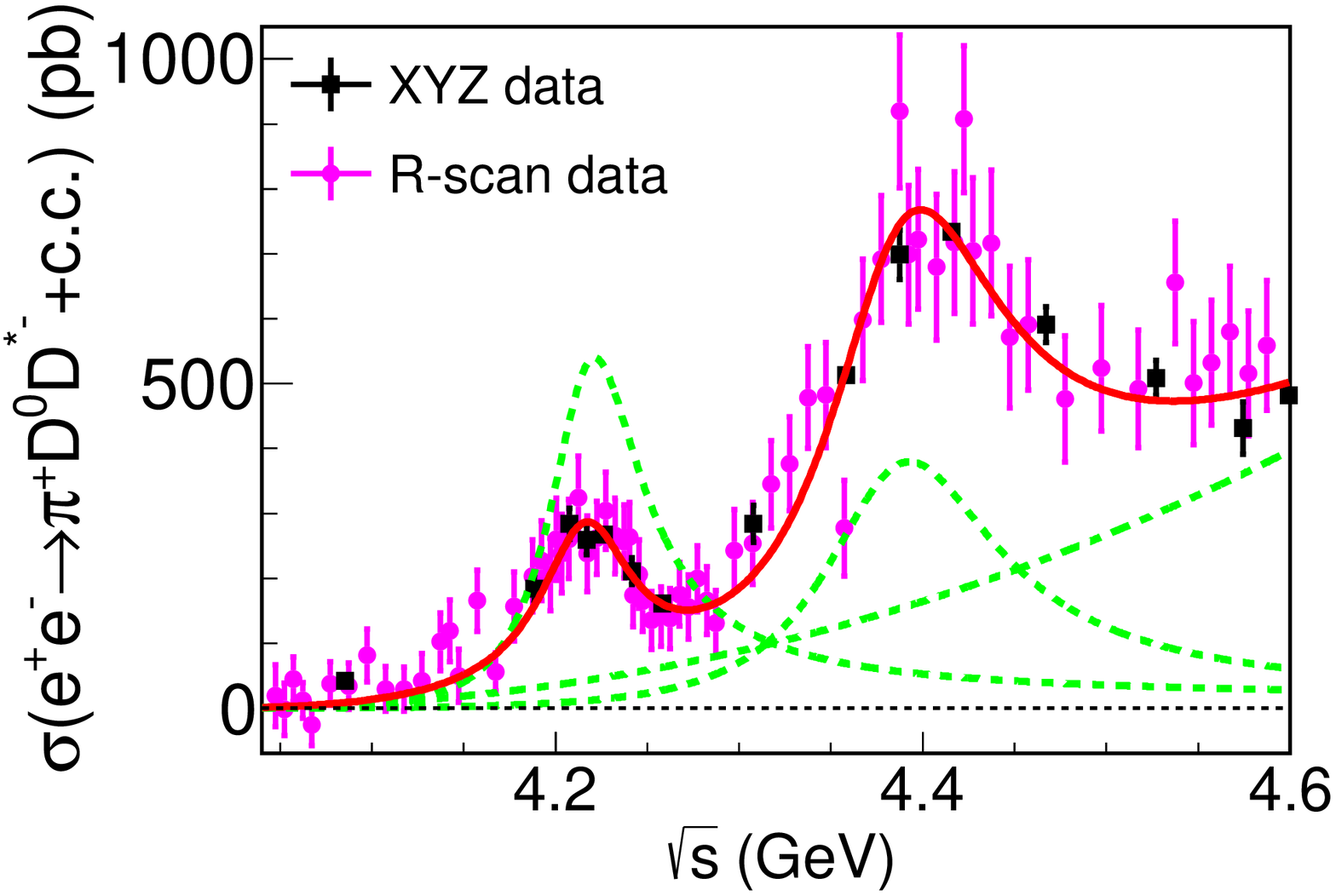}
\includegraphics[width=0.23\textwidth]{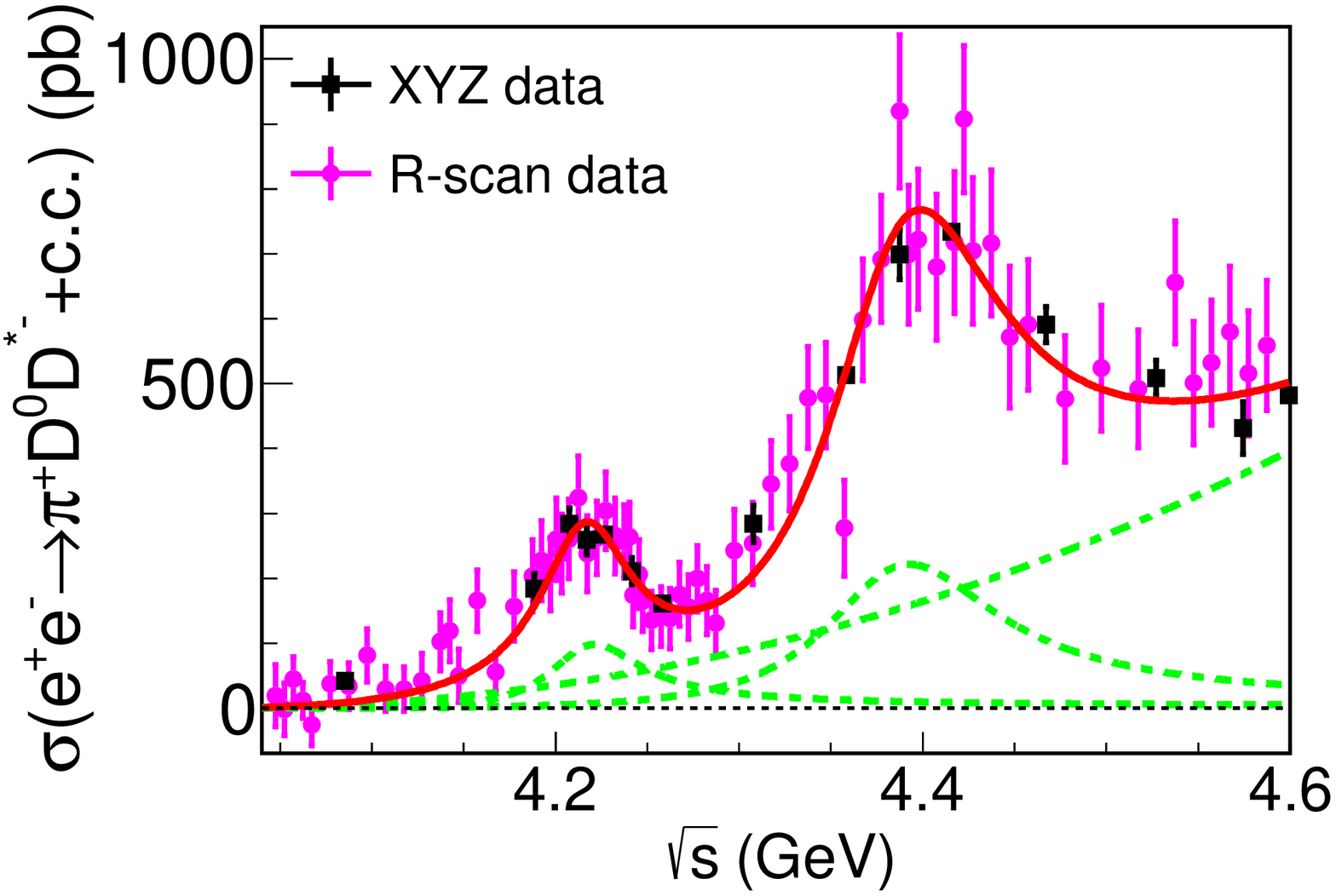}
\includegraphics[width=0.23\textwidth]{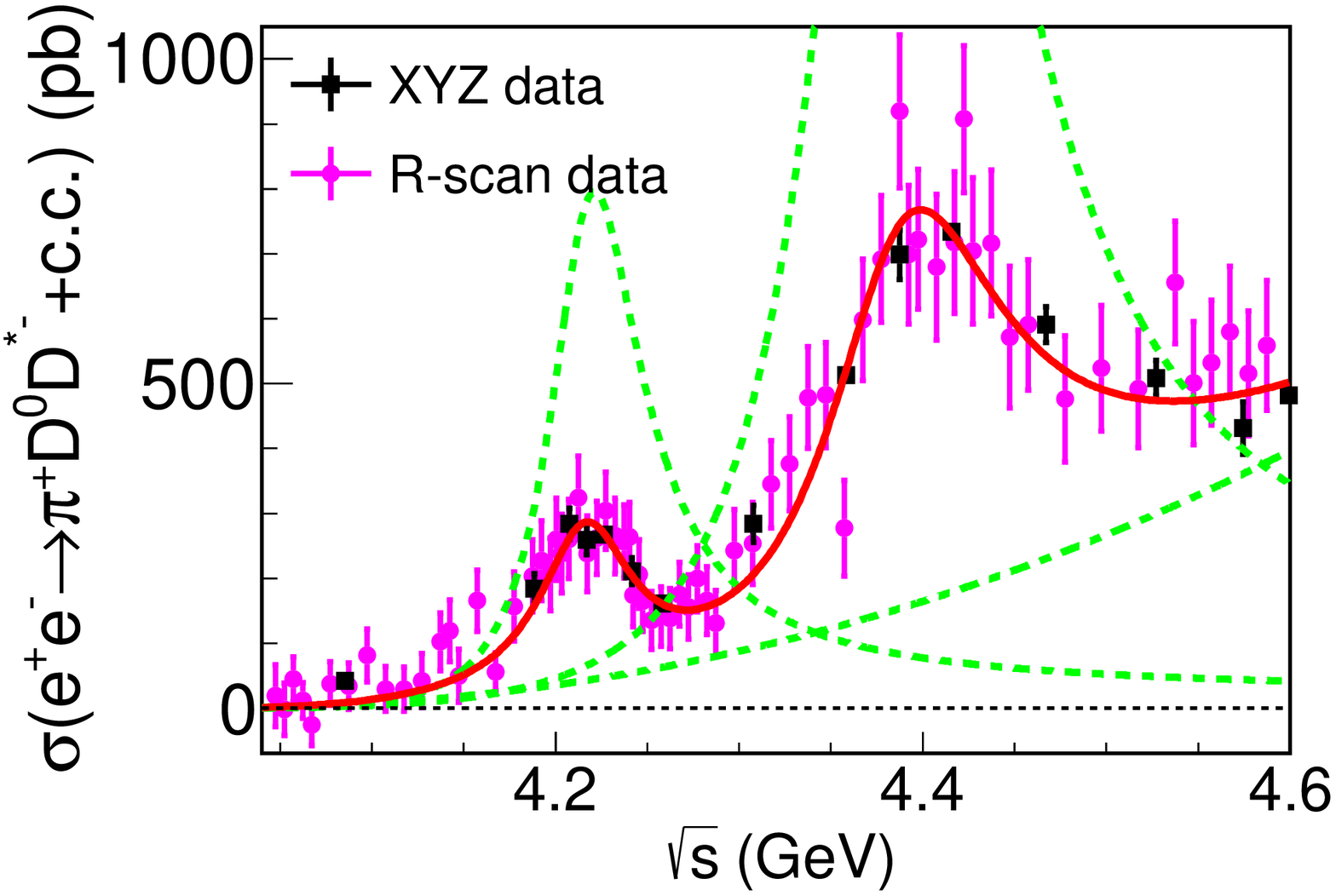}
\includegraphics[width=0.23\textwidth]{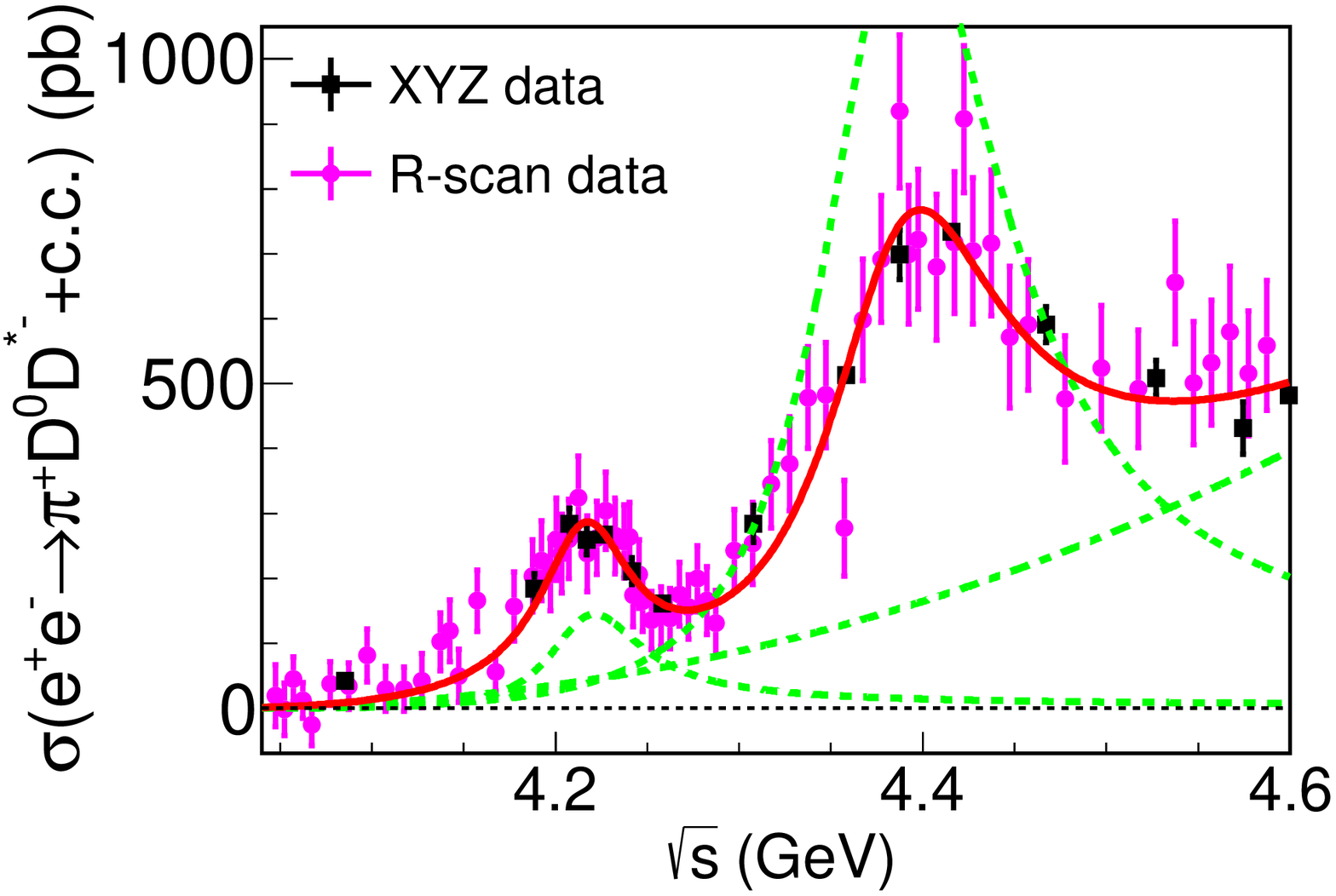}
\caption{The results of the combined fit to the cross sections of $\EE \too \omega\chi_{c0}$, $\pi^{+}\pi^{-}h_c$, $\pi^{+}\pi^{-}J/\psi$, $\pi^{+}\pi^{-}\psi(3686)$ and $\pi^{+}D^{0}D^{*-}+c.c.$ (from the top to the bottom row). The solid red curves show the best fits, and the dashed green ones are individual components.}
\label{fig:fitcrosssection}
\end{center}
\end{figure*}

\begin{table*}[htbp]
\begin{center}
\caption{ The fitted parameters from the combined fit to the cross sections of $\EE \too \omega\chi_{c0}$, $\pi^{+}\pi^{-}h_c$, $\pi^{+}\pi^{-}J/\psi$, $\pi^{+}\pi^{-}\psi(3686)$ and $\pi^{+}D^{0}D^{*-}+c.c.$ The first uncertainties are statistical, and the second systematic.}
\label{tab:fitresult}
\begin{tabular}{cccccc}
  \hline
  \hline
  \quad \quad \qquad Parameter \qquad \qquad \qquad & \quad \quad \qquad $Y(4220)$ \qquad \qquad \qquad & \quad \quad \qquad $Y(4390)$ \qquad \qquad \qquad & \quad \quad \qquad $Y(4660)$ \qquad \qquad \qquad \\
  \hline
  $M$ (MeV/$c^{2}$) & $4216.5\pm1.4\pm3.2$ & $4383.5\pm1.9\pm6.0$ & $4623.4\pm10.5\pm16.1$  \\
  $\Gamma$ (MeV) & $61.1\pm2.3\pm3.1$ & $114.5\pm5.4\pm9.9$ & $106.1\pm16.2\pm17.5$  \\
  \hline
  \hline
\end{tabular}
\begin{tabular}{ccccc}
  \\
  \hline
  \hline
  \quad \quad Parameter \qquad \qquad & \quad \quad SolutionI \qquad \qquad & \quad \quad SolutionII \qquad \qquad & \quad \quad SolutionIII \qquad \qquad & \quad \quad SolutionIV \qquad \qquad   \\
  \hline
  $\Gamma^{Y(4220)}_{e^+e^-}\mathcal{B}(Y(4220) \too \omega\chi_{c0})$ (eV) & $3.5\pm0.4\pm0.5$ & & &  \\
  \hline
  $\Gamma^{Y(4220)}_{e^+e^-}\mathcal{B}(Y(4220) \too \pi^{+}\pi^{-}h_{c})$ (eV) & $6.5\pm0.5\pm1.1$ & $3.1\pm0.2\pm0.8$ &  &  \\
  $\Gamma^{Y(4390)}_{e^+e^-}\mathcal{B}(Y(4390) \too \pi^{+}\pi^{-}h_{c})$ (eV) & $15.1\pm1.0\pm2.8$ & $7.5\pm0.6\pm1.8$ &  &  \\
  \hline
  $\Gamma^{Y(4220)}_{e^+e^-}\mathcal{B}(Y(4220) \too \pi^{+}\pi^{-}J/\psi)$ (eV) & $10.5\pm0.5\pm1.7$ & $12.3\pm0.7\pm2.1$ & $3.7\pm0.3\pm0.6$ & $3.1\pm0.3\pm0.6$ \\
  $\Gamma^{Y(4390)}_{e^+e^-}\mathcal{B}(Y(4390) \too \pi^{+}\pi^{-}J/\psi)$ (eV) & $0.3\pm0.1\pm0.1$ & $12.1\pm0.7\pm3.2$ & $10.4\pm0.6\pm2.3$ & $0.3\pm0.1\pm0.1$ \\
  \hline
  $\Gamma^{Y(4220)}_{e^+e^-}\mathcal{B}(Y(4220) \too \pi^{+}\pi^{-}\psi(3686))$ (eV) & $1.6\pm0.3\pm0.3$ & $1.5\pm0.3\pm0.3$ & $1.6\pm0.3\pm0.3$ & $1.5\pm0.3\pm0.3$ \\
  $\Gamma^{Y(4390)}_{e^+e^-}\mathcal{B}(Y(4390) \too \pi^{+}\pi^{-}\psi(3686))$ (eV) & $13.4\pm1.1\pm1.4$ & $9.9\pm1.0\pm1.2$ & $13.4\pm1.1\pm1.4$ & $9.9\pm1.0\pm1.2$ \\
  $\Gamma^{Y(4660)}_{e^+e^-}\mathcal{B}(Y(4660) \too \pi^{+}\pi^{-}\psi(3686))$ (eV) & $8.8\pm1.2\pm1.4$ & $3.0\pm0.5\pm0.6$ & $8.8\pm1.2\pm1.4$ & $3.0\pm0.5\pm0.6$ \\
  \hline
  $\Gamma^{Y(4220)}_{e^+e^-}\mathcal{B}(Y(4220) \too \pi^{+}D^{0}D^{*-}+c.c.)$ (eV) & $39.0\pm2.5\pm3.1$ & $7.1\pm0.6\pm1.3$ & $57.5\pm3.0\pm6.1$ & $10.5\pm1.1\pm2.7$ \\
  $\Gamma^{Y(4390)}_{e^+e^-}\mathcal{B}(Y(4390) \too \pi^{+}D^{0}D^{*-}+c.c.)$ (eV) & $55.4\pm5.7\pm7.8$ & $32.4\pm2.1\pm2.8$ & $313.6\pm13.9\pm26.4$ & $183.1\pm11.2\pm19.3$ \\
  \hline
  \hline
\end{tabular}
\end{center}
\end{table*}

We fit to the cross sections of $\EE \too \omega\chi_{c0}$, $\pi^{+}\pi^{-}h_c$, $\pi^{+}\pi^{-}J/\psi$, $\pi^{+}\pi^{-}\psi(3686)$ and $\pi^{+}D^{0}D^{*-}+c.c.$ simultaneously. The fits for $\EE \too \omega\chi_{c0}$, $\pi^{+}\pi^{-}h_c$, $\pi^{+}\pi^{-}J/\psi$, $\pi^{+}\pi^{-}\psi(3686)$ and $\pi^{+}D^{0}D^{*-}+c.c.$ are found to have one solution, two solutions, four solutions, four solutions and four solutions with the same minimum values of $\chi^2$, respectively. The masses and widths of the resonances are identical, but the $\Gamma_{e^+e^-}\mathcal{B}_f$ vary with the different solutions for each process.

Figure~\ref{fig:fitcrosssection} shows the fit results with a goodness of the fit is $\chi^{2}/ndf=460/474=0.97$, corresponding to a confidence level of $67\%$. The good fit indicates that the assumption that the two resonances $Y(4220)$ and $Y(4390)$ are same two states in these processes is reasonable. From fit results, we can get $M_{Y(4220)}=(4216.5\pm1.4)$ MeV/$c^{2}$, $\Gamma_{Y(4220)}=(61.1\pm2.3)$ MeV; $M_{Y(4390)}=(4383.5\pm1.9)$ MeV/$c^{2}$, $\Gamma_{Y(4390)}=(114.5\pm5.4)$ MeV; $M_{Y(4660)}=(4623.4\pm10.5)$ MeV/$c^{2}$, $\Gamma_{Y(4660)}=(106.1\pm16.2)$ MeV. The all obtained resonant parameters from fit are listed in Table~\ref{tab:fitresult}.

From the fit results, the obtained parameters of $Y(4660)$ are quite different from Belle's results~\cite{pipipsip-belle}. There are two main reasons. One is that the interference between $Y(4390)$ and $Y(4660)$ has large influence on $Y(4660)$'s parameters. We can see the obtained combined $Y(4390)$'s parameters are very different from the $Y(4360)$'s parameters from Belle's results, it will lead to the $Y(4660)$'s parameters are also different. Another is that the data point at 4.6 GeV from BESIII has very small error, so the fitted $Y(4660)$'s BW curve is influenced greatly by this data point. From Fig.~\ref{fig:fitcrosssection}, we can see that in order to cover the data point, the $Y(4660)$'s BW curve has to have some deviations from Belle data points around 4.66 GeV.

The systematic uncertainties on the resonant parameters in the combined fit to the cross sections of $\EE \too \omega\chi_{c0}$, $\pi^{+}\pi^{-}h_c$, $\pi^{+}\pi^{-}J/\psi$, $\pi^{+}\pi^{-}\psi(3686)$ and $\pi^{+}D^{0}D^{*-}+c.c.$ are mainly from the uncertainties of the center-of-mass energy determination, parametrization of the BW function, background shape and the cross section measurements.

Since the uncertainty of the beam energy is about 0.8 MeV at BESIII, so the uncertainty of the resonant parameters caused by the beam energy is estimated by varing $\sqrt{s}$ within 0.8 MeV for BESIII data. Instead of using a constant total width, we assume an energy dependent width to estimate the uncertainty due to parametrization of BW function. To model the $\EE \too \pi^{+}\pi^{-}J/\psi$ cross section near 4 GeV, a BW function is used to replace the exponential function, and the difference of the fit results in the two methods are taken as the uncertainty from background shape. The uncertainty of the cross section measurements will affect the resonant parameters in fit, we vary the cross sections within the systematic uncertainty, and the difference in the final results are taken as the uncertainty. By assuming all these sources of systematic uncertainties are independent, we add them in quadrature. The systematic uncertainty from the parametrization of the BW function for the paramters mass and width is dominant, while the systematic uncertainty from the cross section measurements for the parameter $\Gamma_{e^+e^-}\mathcal{B}_f$ is dominant.

The leptonic decay width for a vector state is an important quantity for discriminating various theoretical models~\cite{lepton1, lepton2, lepton3}. By considering the isospin symmetric modes of the measured channels, we can estimate the lower limits on the leptonic partial width of the $Y(4220)$ and $Y(4390)$ decays. For an isospin-zero charmonium-like state, we expect

$\mathcal{B}(Y \too \pi\pi h_{c})=\frac{3}{2}\times\mathcal{B}(Y \too \pi^+\pi^- h_{c})$,

$\mathcal{B}(Y \too \pi\pi J/\psi)=\frac{3}{2}\times\mathcal{B}(Y \too \pi^+\pi^- J/\psi)$,

$\mathcal{B}(Y \too \pi\pi \psi(3686))=\frac{3}{2}\times\mathcal{B}(Y \too \pi^+\pi^- \psi(3686))$,

$\mathcal{B}(Y \too \pi D\bar{D}^*)=3\times\mathcal{B}(Y \too \pi^{+}D^{0}D^{*-}+c.c.)$,

\noindent so we have

$\Gamma^{Y(4220)}_{e^+e^-}=\sum\limits_{f}\mathcal{B}(Y(4220) \too f)\times\Gamma^{Y(4220)}_{e^+e^-}$

$\qquad \qquad = \mathcal{B}(Y(4220) \too \omega\chi_{c0})\times\Gamma^{Y(4220)}_{e^+e^-} +$

$\qquad \qquad \quad \mathcal{B}(Y(4220) \too \pi\pi h_{c})\times\Gamma^{Y(4220)}_{e^+e^-} +$

$\qquad \qquad \quad \mathcal{B}(Y(4220) \too \pi\pi J/\psi)\times\Gamma^{Y(4220)}_{e^+e^-} +$

$\qquad \qquad \quad \mathcal{B}(Y(4220) \too \pi\pi \psi(3686))\times\Gamma^{Y(4220)}_{e^+e^-} +$

$\qquad \qquad \quad \mathcal{B}(Y(4220) \too \pi D\bar{D}^*)\times\Gamma^{Y(4220)}_{e^+e^-} + \cdots$

and

$\Gamma^{Y(4390)}_{e^+e^-}=\sum\limits_{f}\mathcal{B}(Y(4390) \too f)\times\Gamma^{Y(4390)}_{e^+e^-}$

$\qquad \qquad = \mathcal{B}(Y(4390) \too \pi\pi h_{c})\times\Gamma^{Y(4390)}_{e^+e^-} +$

$\qquad \qquad \quad \mathcal{B}(Y(4390) \too \pi\pi J/\psi)\times\Gamma^{Y(4390)}_{e^+e^-} +$

$\qquad \qquad \quad \mathcal{B}(Y(4390) \too \pi\pi \psi(3686))\times\Gamma^{Y(4390)}_{e^+e^-} +$

$\qquad \qquad \quad \mathcal{B}(Y(4390) \too \pi D\bar{D}^*)\times\Gamma^{Y(4390)}_{e^+e^-} + \cdots$

By inserting the numbers from Table~\ref{tab:fitresult}, considering the solutions with the smallest $\mathcal{B}(Y(4220) \too f)\times\Gamma^{Y(4220)}_{e^+e^-}$ and $\mathcal{B}(Y(4390) \too f)\times\Gamma^{Y(4390)}_{e^+e^-}$, we obtain

$\Gamma^{Y(4220)}_{e^+e^-}=(3.5\pm0.4\pm0.5)+\frac{3}{2}\times(3.1\pm0.2\pm0.8)+$

$\qquad \qquad \quad \frac{3}{2}\times(3.1\pm0.3\pm0.6)+\frac{3}{2}\times(1.5\pm0.3\pm0.3)+$

$\qquad \qquad \quad 3\times(7.1\pm0.6\pm1.3) + \cdots$ eV

$\qquad \qquad = (36.4\pm2.0\pm4.2) + \cdots$ eV

$\qquad \qquad > (36.4\pm2.0\pm4.2)$ eV, \\

and

$\Gamma^{Y(4390)}_{e^+e^-}=\frac{3}{2}\times(7.5\pm0.6\pm1.8)+\frac{3}{2}\times(0.3\pm0.1\pm0.1)+$

$\qquad \qquad \quad \frac{3}{2}\times(9.9\pm1.0\pm1.2)+3\times(32.4\pm2.1\pm2.8)+$

$\qquad \qquad \quad \cdots$ eV

$\qquad \qquad = (123.8\pm6.5\pm9.0) + \cdots$ eV

$\qquad \qquad > (123.8\pm6.5\pm9.0)$ eV, \\
where the first uncertainties are statistical, and the second systematic.

On the other hand, if we take the results with the largest $\mathcal{B}(Y(4220) \too f)\times\Gamma^{Y(4220)}_{e^+e^-}$ and $\mathcal{B}(Y(4390) \too f)\times\Gamma^{Y(4390)}_{e^+e^-}$ in Table~\ref{tab:fitresult}, we obtain $\Gamma^{Y(4220)}_{e^+e^-} = (206.6\pm9.1\pm18.7) + \cdots$ and $\Gamma^{Y(4390)}_{e^+e^-} = (1001.7\pm41.8\pm79.5) + \cdots$ eV. This means that the leptonic partial widths of $Y(4220)$ and $Y(4390)$ can be as large as $200$ and $1000$ eV or even higher based on current information, because maybe there are some other decay channels for $Y(4220)$ and $Y(4390)$ that we have not observed.

In summary, a combined fit is performed to the cross sections of $\EE \too \omega\chi_{c0}$, $\pi^{+}\pi^{-}h_c$, $\pi^{+}\pi^{-}J/\psi$, $\pi^{+}\pi^{-}\psi(3686)$ and $\pi^{+}D^{0}D^{*-}+c.c.$ by using three resonances $Y(4220)$, $Y(4390)$ and $Y(4660)$. The parameters are determined to be $M_{Y(4220)}=(4216.5\pm1.4\pm3.2)$ MeV/$c^{2}$, $\Gamma_{Y(4220)}=(61.1\pm2.3\pm3.1)$ MeV; $M_{Y(4390)}=(4383.5\pm1.9\pm6.0)$ MeV/$c^{2}$, $\Gamma_{Y(4390)}=(114.5\pm5.4\pm9.9)$ MeV; $M_{Y(4660)}=(4623.4\pm10.5\pm16.1)$ MeV/$c^{2}$, $\Gamma_{Y(4660)}=(106.1\pm16.2\pm17.5)$ MeV, where the first uncertainties are statistical and the second systematic. We emphasize that two resonances $Y(4220)$ and $Y(4390)$ are sufficient to explain these cross sections below 4.6 GeV. The resonances $Y(4320)$, $Y(4360)$ and $Y(4390)$ should be one state. The lower limits of $Y(4220)$ and $Y(4390)$'s leptonic decay widths are also determined to be $(36.4\pm2.0\pm4.2)$ and $(123.8\pm6.5\pm9.0)$ eV. These results will be useful in understanding the nature of charmonium-like states in this energy region. Higher precision measurements around this energy region are desired, this can be achieved in BESIII and BelleII experiments in the further.

\section*{Acknowledgement}
This work is supported by the Foundation of Henan Educational Committee (No. 19A140015), Nanhu Scholars Program for Young Scholars of Xinyang Normal University, and Open Research Program of Large Research Infrastructures (2017), Chinese Academy of Sciences.

\section*{Data Availability}
All the data used in this work are from Ref.~\cite{pipijpsi-bes, omegachic, omegachic2, pipihc-bes, pipihc-cleo, pipijpsi-belle, pipijpsi-babar, pipipsip-belle, pipipsip-babar, pipipsip-bes, piDDstar}.

\end{document}